\documentclass{aa}
\usepackage{graphicx}
\usepackage{psfig}
\usepackage[authoryear]{natbib}
\bibpunct{(}{)}{;}{a}{}{,} 
\newcommand{\targ}{4U~1543--624}
\newcommand{\feka}{Fe $\mathrm{K}_\alpha$}
\begin{document}

   \title{X-ray properties of \targ}

   \author{J. Schultz\thanks{Juho.Schultz@astro.helsinki.fi}} 

   \institute{Observatory,
   P. O. Box 14, FIN-00014 University of Helsinki, Finland}

   \date{Received / Accepted}

\abstract{
\targ~ is a relatively bright persistent low-mass X-ray
binary. Analysis of archival data from ASCA, SAX and RXTE is presented.
The X-ray continuum be can modeled with the standard low-mass X-ray
binary spectrum, an isothermal blackbody and a Comptonized component.
Variations in the luminosity and flux ratio of the continuum components
are seen. An increase in luminosity is accompanied by a decrease
in the blackbody luminosity and a hardening of the spectrum.
Most low-mass X-ray binaries have softer spectra and higher
blackbody luminosities in high luminosity states.
The \feka~ line is seen only in the high luminosity spectra.
A narrow feature near 0.7 keV, previously detected in the
ASCA data, is also seen in the SAX data.
A qualitative model of the system is presented. The X-ray observations
can be explained by a low inclination system (face-on disk)
containing a slowly (P $\gg$ ~ms) rotating neutron star.
A slowly rotating neutron star would imply either that the system is a
young low-mass X-ray binary, or that the accretion rate
is unusually low. The empirical relation between optical and X-ray
luminosity and orbital period suggests a relatively short period.
\keywords{binaries:close -- stars:individual (\targ) -- X-rays:binaries}
}
\maketitle

\section{Introduction}
In low-mass X-ray binaries (LMXB) a stellar-mass black hole or
neutron star accretes from a low-mass companion star.
X-ray emission powered by the accretion flow dominates the
bolometric luminosity in LMXBs. Detailed studies of LMXBs may
be helpful in understanding the physics of accretion and
compact objects, as well as the evolution of close binary stars.

\targ~ is a relatively bright LMXB
discovered by UHURU. It has been observed
at roughly constant flux levels by most major X-ray satellites
over the last decades \citep{Singh, Chr97, Asa00, Jue01}.
The optical counterpart has been identified as a faint
$(B \approx 20)$ star \citep{McC78}. (For a finding chart,
see also Apparao et al. \citep{App78}). Spectral analysis
of EXOSAT data shows that the X-ray continuum can be modeled
with an isothermal blackbody (BB) and a Comptonized component
\citep{Singh}, a model that fits most LMXB spectra quite well
\citep{Whi88}. Narrow spectral features have also been detected:
the \feka~ line at $\simeq 6.5 \, \mathrm{keV}$ \citep{Singh, Got95, Asa00} 
and a feature at $\simeq 0.7 \, \mathrm{keV}$, which 
may be an emission line or an artifact caused by enhanced
Ne absorption \citep{Jue01}. In this paper, archival
SAX, ASCA and RXTE observations of \targ~ are analyzed.
Results of temporal and spectral variability analysis
are discussed.

\section{Observations and data reductions}

\begin{table}
\caption[]{List of pointed observations used for detailed spectral
analysis. Exposure times are in kiloseconds. \label{obstable}}
\begin{tabular}{llll}
\hline
Satellite & Date & Instrument & Exposure \\
\hline
ASCA & 17/8/1995   & GIS  & 12 \\
     &             & SIS  & 10 \\
SAX  & 21/2/1997   & LECS & 7  \\
     &             & MECS & 18 \\
SAX  & 1/4/1997    & LECS & 5  \\
     &             & MECS & 18 \\
XTE  & 5/5/1997    & PCA  & 3  \\
XTE  & 6/5/1997    & PCA  & 8  \\
XTE  & 7/5/1997    & PCA  & 6  \\
XTE  & 12/5/1997   & PCA  & 7  \\
XTE  & 14/5/1997   & PCA  & 3  \\
XTE  & 22/9/1997   & PCA  & 5  \\
XTE  & 13/10/1997  & PCA  & 5  \\
\hline
\end{tabular}

\end{table}

Observations made with the narrow-field
instruments of SAX, i.e. LECS \citep{LECS}, MECS \citep{MECS}, 
HPGSPC \citep{HPGSPC} and PDS \citep{PDS},  
all three RXTE \citep{XTE} instruments, i.e. HEXTE, PCA \citep{PCA} 
and ASM \citep{ASM},
and both ASCA \citep{ASCA} instruments, i.e. GIS and SIS are analyzed.
Of these, LECS, MECS, PCA, GIS and SIS 
(See Table \ref{obstable}) had data with sufficient
S/N for spectral analysis.

The event lists provided by on-line archives
were used for both SAX and ASCA data.
Raw RXTE data were used for both PCA
(`Good Xenon' or `Standard 2'-modes) and HEXTE analysis.
The `definitive' RXTE ASM data were used to 
study the long-term variability of \targ.

\subsection{SAX}

For the SAX data, cleaned event lists were downloaded
from the ASDC website. The LECS and MECS spectra were
extracted from a 4' region at the center of the
field-of-view. The standard response files provided
by ASDC were used. The background spectra were extracted from the
blank-sky event lists available at ASDC using the same extraction
regions as for the source spectra. The PDS and HPGSPC spectra
were background-dominated, and not used for further analysis.

\subsection{ASCA}
The standard filtering criteria were used to produce a cleaned
event list from the raw event list.
The GIS2 count rate was $\sim 16 \, \mathrm{cps}$, so significant pileup
is expected in the SIS data. As the SIS data was in single-frame mode,
the spectrum could be extracted with the {\tt corpileup} tool.
For SIS, the background spectrum was extracted from a blank-sky event list.
The background was compared to a spectrum extracted from the science
data from a region without sources at the same off-axis distance
as the source. The difference between alternative background models
was small, and as the SIS data is affected by pileup and the
background flux is less than 0.5 \% of the source flux, instrumental
effects probably dominate the uncertainties of the data. 
For GIS, a blank-sky event list was used for background extraction.
The extraction radii used were 6' for GIS and 4' for SIS.
The response files were generated by the techniques recommended
in the  ASCA Data Reduction
Guide\footnote{http://heasarc.gsfc.nasa.gov/docs/asca/abc/abc.html}.
To improve the signal-to-noise ratio, the separate
data and calibration files of the two GIS and two SIS detectors
were combined into files containing all GIS data and all SIS data.

\begin{figure}
\resizebox{\hsize}{!}{\psfig{figure=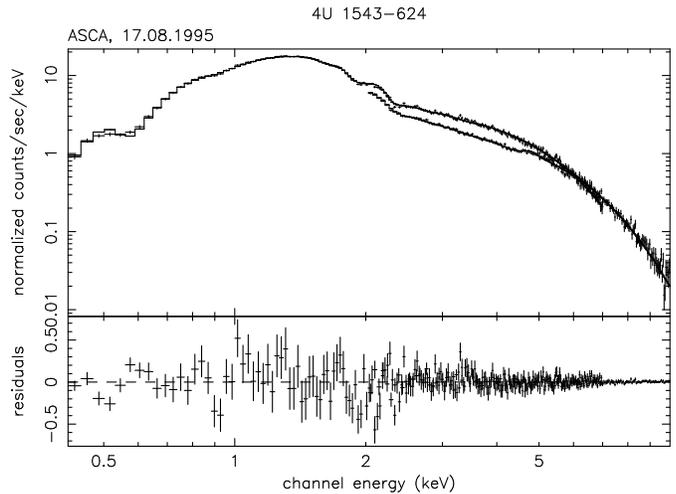,angle=270}}
      \caption{The ASCA data (SIS and GIS), a fitted spectral model
      with blackbody and Comptonized components, and line emission 
      at $0.676 \, \mathrm{keV}$. The lower panel
      shows the residuals.}
      \label{ASCA}
\end{figure}

\subsection{RXTE}
PCA and HEXTE target and background spectra and lightcurves
were extracted from the raw data. 
The PCA data was also used to create response files
and power spectra.
The data were processed as recommended in the RXTE Users'
Guide\footnote{http://heasarc.gsfc.nasa.gov/docs/xte/abc/contents.html}
to get cleaned event lists.  For HEXTE, the responses available
at the HEXTE calibration status 
website\footnote{http://mamacass.ucsd.edu:8080/hexte/hexte\_calib.html},
were used. Fitting a powerlaw to the summed  HEXTE 
spectra ( between $15$--$200 \, \mathrm{keV}$) gives a
$3\sigma$ upper limit of $1.4 \cdot 10^{-13} \, \mathrm{W} \mathrm{m}^{-2}$
for the high-energy flux ($15$--$200 \, \mathrm{keV}$ band).
No further spectral analysis of the HEXTE data was made.

\subsection{Spectral fitting}
The XSPEC v. 11.0 spectral fitting package \citep{Arn96}
was used for spectral analysis. The data used for fitting were in the
$0.14$--$4 \, \mathrm{keV}$ band for LECS, $1.8$--$10 \, \mathrm{keV}$ 
for MECS, $0.4$--$7 \, \mathrm{keV}$ for SIS, 
$1.0$--$10 \, \mathrm{keV}$ for GIS
and $3$--$20 \, \mathrm{keV}$ for PCA. 

To avoid possible instrument inter-calibration 
problems in the cases of ASCA (GIS/SIS) and SAX (MECS/LECS),
a constant multiplier for one of the instruments was included in
the fits. For ASCA, the difference in flux between GIS and SIS was
$2.7 \pm 0.8 \, \%$. In the SAX data, the MECS/LECS ratio was $0.70 \pm
0.01$ for the final spectral model. The ratio should be in
the range $0.7$--$1.0$ \citep{Fio99},
so the result is on the edge of the allowed range. 

A systematic error of 1 \% is usually added to the PCA data
for spectral fitting. This was not applied,
as the best models have $\chi_\nu \simeq 1$ even without
systematic errors. To justify the inclusion of additional
spectral components, the $\Delta \chi^2$ was also calculated
with systematic errors of 1 \% and 2 \%.
During the fits to PCA data, the column density of interstellar matter
was held fixed to the weighted mean of 
the values derived from ASCA and SAX fits,
$N_{\mathrm{H}} = 33.0 \pm 3.5 \cdot 10^{24} \, \mathrm{m} ^{-2}$.
As no PCA data below $3\, \mathrm{keV}$ was used in the analysis, 
column depths of this order should have only minor effects.
To justify the fixed column depth,
fits with a freely varying column depth were made to the PCA spectra.
The column densities tended to diverge toward zero.
Upper limits for the $N_{\mathrm{H}}$ could not always be derived.
When the limits could be derived, typical $3\sigma$ upper limits were
of the order $1-3  \cdot 10^{26} \, \mathrm{m} ^{-2}$.
The other spectral parameters were nearly identical,
but their error ranges increased by about 20 \%.
The spectral fits to PCA data give flux estimates of the order
$3 \cdot 10^{-14} \, \mathrm{W} \mathrm{m}^{-2}$ 
for the flux above $15 \, \mathrm{keV}$, which is
consistent with the HEXTE data.

Several single component spectral models (isothermal and 
multicolor blackbody \citep{Mak86}, thermal bremsstrahlung, 
powerlaw, cut-off powerlaw and some Comptonization
models) were tried, all with photoelectric absorption from the ISM.
No single component model gave an acceptable fit,
with typical values of $\chi_\nu^2 > 2.5$.
Two-component spectra with an isothermal blackbody (BB) as one component and
a cut-off powerlaw or one of the Comptonization models, CompST 
\citep{Sun80} or CompTT \citep{Tit94}, as the other component
gave the best fits. The BB-CompST fit was in most
cases the best. In three cases, CompTT or powerlaw gave a
better fit than CompST, but the differences were not statistically 
significant. The improvement in $\chi_\nu^2$ was less than 0.01.
The high optical depths of the Comptonization fits mean
that the fitted spectra are relatively close to a cut-off powerlaw.
In two cases, a combination of a multicolor blackbody and a
powerlaw also gave a statistically acceptable fit.
The CompST model was adopted for further analysis,
as this would allow a meaningful comparison between
the spectra of different observations. 

The continuum fits were further improved by adding gaussian features
when the residuals of the fits seemed to have a line component.
In the SAX and ASCA data, a gaussian feature was detected near
$0.67 \, \mathrm{keV}$. The \feka~ line could not be detected in
SAX and ASCA data, but an upper limit was derived for the line flux.
This was done by adding a gaussian feature to the fit. The
centroid energy was allowed to vary in the $6.3$--$7.1 \, \mathrm{keV}$
range. As the fit failed to find a statistically significant line, the
$3\sigma$ upper limit of the line flux was
estimated using the `error' command of XSPEC.
For the RXTE/PCA data, the same procedure detected the \feka~ lines,
with equivalent widths of $\sim 80$--$160  \, \mathrm{eV}$.
The improvement in $\chi^2$ was of the order 200 
(for 37 degrees of freedom). No systematic errors were used
in the fitting, but the 
\feka~ line is statistically significant even with systematic errors.
A systematic error of 1 \% produces $\Delta \chi^2 \approx 60$ and
a 2 \% error $\Delta \chi^2 \approx 30$ for adding the \feka~ line.
The upper limit of the ASCA line flux is of the order 10 \% of the
detected RXTE/PCA flux. The parameters describing the
final spectral fits are listed in Table \ref{conttable}
(continuum parameters) and \ref{linetable} (line parameters).

To quantify the instrumental systematic
effects on the iron line parameters, the RXTE calibration observations
of the Crab were used. The standard products of Crab observations 
made near our observations (27.4, 9.5., 20.7, 12.9, 29.9, 12.10 and
13.10) were analyzed. The PCA $3$-$20\, \mathrm{keV}$
spectra were fitted with an absorbed powerlaw and a gaussian line.
The line centroid energies were near $7\, \mathrm{keV}$.
The fluxes of the line and continuum were calculated in a 1 keV
band centered on the line centroid energy. The highest line to
continuum ratio was 1.14 \%. Typically the statistical error of the
line to continuum ratio was a factor of 2-3, giving a 90 \% upper
limit of 3 \% for the ratio. As the line to continuum
flux ratios of the spectral fits of \targ ~ vary between 3--7\%,
it can be concluded that the iron line has been detected. 
However, the line parameters of the RXTE fits listed in Table \ref{linetable}
are partially produced by instrumental uncertainties and therefore
should be treated with some caution.

\subsection{Temporal variability analysis}

The ASCA/GIS and SAX/MECS  lightcurves were extracted from
cleaned event lists. RXTE (HEXTE and PCA) lightcurves were extracted
from the event lists generated from the raw data. The HEXTE lightcurves
were extracted in the $15$--$200 \, \mathrm{keV}$ band.
For PCA data, three bands were used, $2$--$5 \, \mathrm{keV}$ (soft),
$5$--$13 \, \mathrm{keV}$ (medium) and  $13$--$40 \, \mathrm{keV}$
(hard) in the extraction process. No statistically significant
periodic variations were detected in a Lomb--Scargle
periodogram \citep{Sca82} made from the inidividual lightcurves.
Absence of X-ray modulation suggests a relatively low inclination
of the accretion disk.

Quasi-periodic oscillations (QPOs) were searched in the frequency
range $10 \, \mathrm{kHz}$ --$0.01 \, \mathrm{Hz}$ 
(in the $3$--$10 \, \mathrm{keV}$ band).
Power spectra were extracted separately 
for each good time interval of all RXTE/PCA observations. 
After this, the power spectra were co-added to get a summed
power spectrum. No statistically significant features
(down to a few \% rms amplitude) were seen in either the
individual power spectra or the weighted mean.

\begin{table*}
\caption[]{The continuum parameters of the spectral
fits with 90 \% error estimates. Errors for the parameters 
derived from the fitted parameters have been calculated
according to the propagation of errors.  
Line components are presented in
Table \ref{linetable}.
The values of luminosity ($L$) are unabsorbed 2--10 keV band values.
For RXTE/PCA, the 2--3 keV part is based on extrapolation of the
spectral fit. An assumed
distance of 10 kpc has been used for the luminosity and BB
radius estimates. The $N_{\mathrm{H}}$ values denoted with 
F have been held fixed
(at $33 \cdot 10^{24} \, \mathrm{m}^{-2}$)
during the fits. The Comptonization parameter is
$\gamma_{\mathrm{C}} = 
4 k T_{\mathrm{C}} \tau_{\mathrm{C}}^2 / m_{\mathrm{e}} c^2$. 
No systematic errors are
included in the $\chi^2$ values of the RXTE/PCA data.
\label{conttable}}
\begin{tabular}{lllllllllll}
\hline
Satellite & Date 
& $L$ & $L_{\mathrm{BB}}/L$ 
& $N_{\mathrm{H}}$ & $T_{\mathrm{BB}}$
& $R_{\mathrm{BB}}$ & $T_{\mathrm{C}}$
& $\tau_{\mathrm{C}}$ & $\gamma_{\mathrm{C}}$ 
& $\chi_\nu$ (dof) \\
& &$10^{28} \mathrm{W}$& \% &$10^{24}\mathrm{m}^{-2}$& 
keV & km & 
keV & & & \\
\hline
ASCA & 17.8.95  
& $86$ & $75$ 
& $30_{-4}^{+3}$ & $1.55_{-0.03}^{+0.03}$
& $3.21_{-0.11}^{+0.15}$ & $0.54_{-0.03}^{+0.02}$ 
& $33.0_{-2.8}^{+4.4}$ & $4.6_{-0.8}^{+1.2}$ 
& 1.24 (394)\\

SAX  & 21.2.97  
& $101$ & $73$ 
& $42_{-8}^{+6}$ &$1.677_{-0.017}^{+0.010}$
& $2.99_{-0.09}^{+0.02}$ & $0.70_{-0.03}^{+0.06}$ 
& $22.0_{-0.9}^{+3.0}$ & $2.7_{-0.3}^{+0.8}$ 
& 1.40 (527)\\

SAX  & 1.4.97   
& $88$ & $73$
& $35_{-5}^{+7}$ & $1.575_{-0.008}^{+0.012}$ 
& $3.14_{-0.09}^{+0.09}$ & $0.57_{-0.03}^{+0.04}$ 
& $29.7_{-3.6}^{+4.2}$ & $3.9_{-1.0}^{+1.2}$ 
& 1.24 (528)\\

RXTE  & 5.5.97   
& $123$ & $38$ 
& $33 $ F & $1.47_{-0.05}^{+0.02}$
& $3.06_{-0.08}^{+0.08}$ & $3.11_{-0.22}^{+0.20}$ 
& $10.7_{-0.7}^{+0.9}$ & $2.8_{-0.4}^{+0.5}$ 
& 0.79 (37)\\

RXTE  & 6.5.97   
& $128$ & $40$ 
& $33 $ F & $1.48_{-0.08}^{+0.03}$
& $3.18_{-0.13}^{+0.11}$ & $3.02_{-0.34}^{+0.36}$ 
& $10.9_{-1.3}^{+1.7}$ & $2.8_{-0.7}^{+0.9}$
& 0.92 (37)\\

RXTE  & 7.5.97 
& $123$ & $42$ 
& $33 $ F & $1.51_{-0.05}^{+0.01}$
& $3.05_{-0.10}^{+0.08}$ & $3.18_{-0.29}^{+0.29}$
& $10.4_{-0.9}^{+1.3}$ & $2.7_{-0.5}^{+0.7}$ 
& 1.16 (37)\\

RXTE  & 12.5.97  
& $129$ & $40$ 
& $33 $ F & $1.48_{-0.03}^{+0.04}$
& $3.14_{-0.09}^{+0.07}$ & $3.04_{-0.16}^{+0.23}$ 
& $11.1_{-1.2}^{+0.7}$ & $2.9_{-0.6}^{+0.5}$
& 0.96 (37)\\

RXTE  & 14.5.97  
& $130$ & $37$ 
& $33 $ F & $1.49_{-0.07}^{+0.02}$
& $3.01_{-0.14}^{+0.19}$ & $3.53_{-0.49}^{+0.52}$
& $9.2_{-1.1}^{+1.6}$ & $2.3_{-0.7}^{+0.9}$ 
& 0.74 (37)\\ 

RXTE  & 22.9.97  
& $136$ & $35$ 
& $33 $ F & $1.48_{-0.05}^{+0.02}$
& $3.03_{-0.17}^{+0.12}$ & $2.94_{-0.24}^{+0.21}$ 
& $11.1_{-1.0}^{+1.2}$ & $2.8_{-0.5}^{+0.7}$ 
&  0.98 (37)\\

RXTE  & 13.10.97 
& $127$ & $38$ 
& $33 $ F & $1.46_{-0.05}^{+0.04}$
& $3.16_{-0.11}^{+0.11}$ & $3.12_{-0.12}^{+0.30}$ 
& $10.7_{-1.0}^{+1.1}$ & $2.8_{-0.5}^{+0.6}$ 
& 0.71 (37)\\

\hline
\end{tabular}
\end{table*}

To study the long-term variability, the lightcurve of the RXTE ASM
(up to January 2002) was analyzed.
No periodicities were detected in a Lomb--Scargle
periodogram \citep{Sca82}. In order to see if a trend is
present in the ASM data, a linear model 
$F_{\mathrm{ASM}} = A + Bt$  was fitted to the flux values.
$F_{\mathrm{ASM}}$ is the observed ASM count rate
(counts per second, 1 mCrab = 0.075 cps), $A$ and $B$ are
the fit parameters and $t$ is time in years from JD 2450000.5.
The best fit values are $A = 3.269 \pm 0.0082$
and $B  = -0.1735 \pm 0.0044$, but the improvement over
a constant model in $\chi^2$ is marginal (38887 to 37374).

To justify the fitting of a linear model,
the literature was searched for older flux values of \targ.
MIR/Kvant observations \citep{Eme00} on 30.1.1989
show a flux of $65 \pm 11$ mCrab
($1.71 \pm 0.29 \cdot 10^{-12} \, \mathrm{W} \, \mathrm{m}^{-2}$)
in the $2$--$30 \, \mathrm{keV}$ band.
An extrapolation of the linear fit predicts a flux
$58.7 \pm 0.1$ mCrab for the Kvant observation, which is
surprisingly close to the observed value. However, older observations
show that the Kvant value is probably an isolated event and does not
reflect an underlying trend.
All the values listed below are absorbed fluxes in the
$2$--$10 \, \mathrm{keV}$ band, unless otherwise stated. 
The HEAO--1 \citep{Woo84} flux is
$6.58 \pm 0.07 \cdot 10^{-13} \, \mathrm{W} \, \mathrm{m}^{-2}$
(15.8.1977 to 15.2.1978),
Ariel V \citep{War81} 
$6.9 \pm 0.07 \cdot 10^{-13} \, \mathrm{W} \,\mathrm{m}^{-2}$,
(15.10.1974 to 14.3.1980),
varying between $3.0$ --
$12.0 \cdot 10^{-13} \, \mathrm{W} \,\mathrm{m}^{-2}$,
OSO--7 \citep{Mar79} 
$8.09 \pm 0.85 \cdot 10^{-13} \, \mathrm{W} \, \mathrm{m}^{-2}$
(October 1971 to May 1973)
and UHURU \citep{For78} 
$4.58 \pm 0.26 \cdot 10^{-13} \, \mathrm{W} \, \mathrm{m}^{-2}$
(12.12.1970 to 18.3.1973).
Einstein observations on 17.3.1979 \citep{Chr97} give a flux of
\mbox{$\sim 7.0 \cdot 10^{-13} \, \mathrm{W} \, \mathrm{m}^{-2}$}
in the $0.5$--$20 \, \mathrm{keV}$ band. 
The fluxes of the older surveys are clearly inconsistent with
extrapolations of the linear model. The flux estimates of the 1970's
missions have been derived from multiple scans, and 
more likely represent the average flux. 
Combining the results of the older missions with the
ASM monitoring data indicates that the source variations
are mainly irregular. The flux differences of the observations
listed above, as well as the results of the pointed observations
discussed below, indicate that flux variations by a factor of 2 may
be seen on timescales of a few weeks.

\section{Spectroscopic results}

\subsection{Continuum}

The best-fit model has the same continuum components,
an isothermal blackbody and a Comptonized component,  as
the fit to older EXOSAT data \citep{Singh}, and
the BB and line parameters also have values consistent with
the EXOSAT fit. The high S/N of the more modern instruments
naturally gives much better constraints on the model parameters.
The parameters of the Comptonized component are different
(however, these were only weakly constrained for the EXOSAT fit,
which has $kT_{\mathrm{C}} > 4 \, \mathrm{keV}$ and $\tau_{\mathrm{C}} < 4.8$).
All continuum parameters are given
in Table  \ref{conttable}. 
The absorbing column depth is about one order of magnitude
lower in the fits to the new data.
In order to study this discrepancy, fits to
archival EXOSAT data were made, with a column depth fixed to 
$N_H = 33 \cdot 10^{24} \, \mathrm{m} ^{-2}$ (the value derived from
fits to SAX and ASCA data). The values of the spectral parameters
were similar to those derived from the ASCA and SAX fits, but
the reduced chi-squared of the fits was of the order 
$\chi_\nu = 1.4$. The old EXOSAT analysis is significantly better,
$\chi_\nu = 1.16$ \citep{Singh}. The EXOSAT spectrum contains no
information on fluxes below $\sim 1 \, \mathrm{keV}$
and the spectral resolution is rather modest at the
low energy end. Therefore, it is possible that the absorbing
column is not strongly constrained by the EXOSAT data.

A literature value of
$N_H = 29.9 \pm 0.8 \cdot 10^{24} \, \mathrm{m}^{-2}$
can be found for Einstein data \citep{Chr97}.
Analysis of the ASCA data
with different absorption models gives column densities of
$26$--$37 \cdot 10^{24} \, \mathrm{m}^{-2}$ \citep{Jue01}.
The galactic hydrogen column in the direction of \targ~
is  $N_H \approx 30 \cdot 10^{24} \, \mathrm{m}^{-2}$
\citep{Dic90}. As the absorbing column is constrained better with observations
covering lower energies, the ASCA, SAX and Einstein fits are most
likely representing the correct value. However, the absorption
may be influenced by matter local to \targ~ enriched in
medium--Z elements \citep{Jue01}.

The parameters of the Comptonized component and the overall luminosity
vary significantly between observations. 
This can be interpreted as the system having two distinct states:
when the system has higher luminosity it is referred to as being in
the high state while when it has lower luminosity it is referered to
as being in the low state. In the low state the Comptonized
component is cooler ($kT \simeq 0.6 \, \mathrm{keV}$), with
$\tau \simeq 30$. In the high state the Comptonized component is hotter 
($kT \simeq 3 \, \mathrm{keV}$), with $ \tau \simeq 10$.
The luminosity ($2$--$10 \, \mathrm{keV}$) 
is about 50 \% higher in the high state, and the
spectrum is significantly harder. A difference of this magnitude
can not be explained by calibration errors. 
The BB is marginally hotter in the low state
($1.6$ vs. $1.5  \, \mathrm{keV}$).
The RXTE observations show a high state spectrum.
The ASCA observation and the SAX observation of 1.4.1997 show a
low state spectrum. The SAX observation of 21.2.1997 shows a
low state spectrum but the parameters are a little different from
the two other low state spectra: the overall luminosity is higher, 
the Comptonized component is slightly hotter
and has a lower optical depth.

\subsection{Narrow features}

\begin{table}
\caption[]{Detected Gaussian features. For the 6 keV feature,
upper limits are given for the non-detections. Errors are 90 \%
confidence limits given by the XSPEC 'error' command.
The RXTE fits were made with $N_H$ fixed to $33 \cdot 10^{24} \, \mathrm{m}
^{-2}$). 
\label{linetable}}
\begin{tabular}{llll} 
\hline
Dataset&Centroid&Flux&$\sigma$\\
       &keV&$10^{-14} \, \mathrm{W}\mathrm{m}^{-2}$&keV\\
\hline
ASCA 17.8.95&$0.676 \pm 0.016$&$8.0_{-2.2}^{+4.6}$ 
                                &$0.093_{-.012}^{+.005}$\\
SAX 21.2.97&$0.67 \pm 0.05$&$18_{-12}^{+49}$&$0.11 \pm 0.04$\\
SAX 1.4.97&$0.67 \pm 0.07$&$12_{-4}^{+6}$  &$0.10 \pm 0.03$\\
\hline
ASCA 17.8.95&N/A&$ <0.18 $&N/A\\
SAX 21.2.97&N/A&$ <1.04 $&N/A\\
SAX 1.4.97&N/A&$ <0.31 $&N/A\\

RXTE 5.5.97&$6.67_{-0.10}^{+0.08}$&$1.12_{-0.17}^{+0.41}$ 
                                         &$ 0.75_{-0.06}^{+0.15}$\\

RXTE 6.5.97&$6.65_{-0.17}^{+0.12}$&$0.94_{-0.30}^{+0.61}$ 
                                         &$ 0.63_{-0.25}^{+0.38}$\\

RXTE 7.5.97&$6.56_{-0.16}^{+0.12}$&$0.97_{-0.26}^{+0.41}$ 
                                        &$ 0.78_{-0.16}^{+0.22}$\\

RXTE 12.5.97&$6.61_{-0.14}^{+0.10}$&$0.84_{-0.26}^{+0.33}$ 
                                         &$0.68_{-0.24}^{+0.21}$\\

RXTE 14.5.97&$6.52_{-0.17}^{+0.20}$&$1.89_{-0.68}^{+0.48}$ 
                                         &$0.95_{-0.26}^{+0.21}$\\

RXTE 22.9.97&$6.46_{-0.15}^{+0.11}$&$ 1.43_{-0.36}^{+0.66}$
                                         &$ 0.76_{-0.18}^{+0.22}$\\

RXTE 13.10.97&$6.55_{-0.15}^{+0.12}$&$ 1.23_{-0.31}^{+0.49}$ 
                                          &$ 0.77_{-0.17}^{+0.20}$\\

\hline
\end{tabular}

\end{table}

In the RXTE spectra the \feka ~ line is seen as a gaussian feature
at $\approx 6.5 \, \mathrm{keV}$ with flux 
$\sim  10^{-14} \, \mathrm{W}\, \mathrm{m}^{-2}$. 
Two flux values similar to this,
$1.23 \cdot 10^{-14} \, \mathrm{W}\, \mathrm{m}^{-2}$ \citep{Singh} and
$0.63 \cdot 10^{-14} \, \mathrm{W}\, \mathrm{m}^{-2}$ \citep{Got95} 
have been derived from the EXOSAT observation.
The first value is from a dedicated study, the second one from the
iron line catalogue, so the continuum model may be more refined
in the first fit. A previous analysis of the ASCA
data \citep{Asa00} has given line fluxes of $0.37$ and
$0.59 \cdot 10^{-14} \, \mathrm{W}\, \mathrm{m}^{-2}$
for narrow and broad line models, respectively. However, the
continuum model is fitted to data only in the band
$4$--$10\, \mathrm{keV}$, and the column density used is 
$N_{\mathrm{H}} = 140 \cdot 10^{24} \, \mathrm{m}^{-2}$,
considerably higher than the values derived from the same data.
The possible differences in continuum slope and 
contribution of the iron K absorption edge 
at $7.1 \, \mathrm{keV}$ to the spectrum are most likely
responsible for the difference in the \feka~ line parameters
derived for the EXOSAT and ASCA observations.

\begin{figure}
\resizebox{\hsize}{!}{\psfig{figure=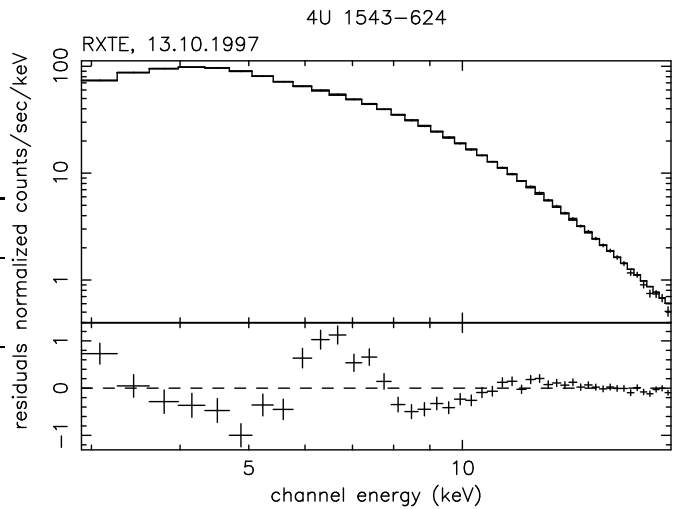,angle=270}}
      \caption{A sample RXTE fit with BB and Comptonized
      component. The residuals indicate a line close to
       $6.5 \, \mathrm{keV}$.}
      \label{PCA_line}
\end{figure}

Non-detection of the \feka~ line in the ASCA and
SAX/MECS spectra suggest variations of at least an order of magnitude
in the line flux. The line is seen in observations
from two different satellites (EXOSAT and RXTE). 
Therefore it is most probably safe to assume that the detection of the line
is not caused by incorrect background subtraction.

Another gaussian feature is seen at $\approx 0.67 \, \mathrm{keV}$
in the SAX/LECS and ASCA/SIS spectra. All three observations have
similar line parameters. The RXTE energy band does not
cover the energy of this line.
The feature may be a neon emission line.
Another possibility is enhanced absorption of
Ne--enriched matter \citep{Jue01}.  The deeper absorption edge
is seen as an artifact that is misinterpreted as a line.

\section{Discussion}

Potential black hole diagnostics include 
ultrasoft X-ray spectra, high-energy powerlaw `tails',
high-soft and low-hard spectral states and millisecond
variability in the hard state \citep{Tan95}. 
As \targ~ exhibits none of these features,
it is assumed that the compact object is a neutron star. 
Asai et al. (2000) label \targ~ as an X-ray burster
(which would confirm the neutron star nature
of the compact object), but no reference is given.

\subsection{The X-ray continuum}

The isothermal blackbody originates from an optically thick
boundary layer between the accretion disk and the neutron star surface
(Sunyaev \& Shakura 1986: see also Inogamov \& Sunyaev 1999).
The Comptonized component is produced by a corona of hot electrons
upscattering soft photons from the disk. The energy of the
Comptonized component is probably derived from the disk
(see e.g. Church et al. (2002) and references therein).

The ratio of the boundary layer and inner disk luminosities
should be close to unity if the neutron star is rotating slowly
\citep{Sun86}. For faster rotators, the boundary layer should be
less luminous, as the velocity of the accreted matter is closer to
that of the neutron star surface.
If the disk is truncated by the neutron star
magnetosphere, its luminosity should be smaller \citep{Whi88}.
Stronger accretion flows should then push the inner disk closer
to the compact object by decreasing the magnetosphere.
The observed BB luminosity of \targ~ is about 3/4 of the total
($L_{\mathrm{BB}}/L_{\mathrm{comp}} \approx 3$) in the low state and
slightly above 1/3 of the total
($L_{\mathrm{BB}}/L_{\mathrm{comp}} \approx 0.5$) in the high state.
The ratio of the continuum component luminosities is relatively close
to theoretical excpectations for a slow rotator,
suggesting the neutron star has not experienced
spin-up to millisecond periods.
Decreasing BB temperature in the high state reduces the
BB luminosity slightly.
It might be that the boundary layer is not radiating as a pure
blackbody in the high state. Part of the luminosity difference
could also be due to smaller inner disk radius, which would 
explain the higher disk (Comptonized component) luminosity. 
The blackbody fraction of the luminosity is clearly higher than
in most LMXBs and in general the BB luminosity correlates
strongly with total luminosity \citep{Chu02, Whi88}.

\subsection{Comptonizing corona}

Observational evidence for Comptonizing coronae is
quite convincing \citep{Whi88}.
No generally accepted model for the geometry
(location, size and shape) of the Comptonizing coronae in LMXBs exists. 
A few geometries are discussed qualitatively below.

Should the corona be close to the boundary layer,
the high optical depth of the Comptonized component would
probably completely block any boundary layer emission.
A clumpy Comptonizing corona could have high optical depth,
and allow some boundary layer emission to shine through.
Only a fraction of the boundary layer emission
would then be intercepted by the corona. Reprocessed
boundary layer photons would probably be seen as an
additional X-ray continuum component. The higher temperature 
and lower optical depth of the clumps in the high state would be 
explained by evaporation of the outer parts of the clumps.

For a face-on system, a toroidal corona where the center
is empty could produce the observed spectra.
The accretion flow would be prevented from entering the central region
by radiation pressure or the magnetic field of the neutron star.
Near-Eddington accretion rates are needed to reach levels
of radiation pressure that can influence the accretion flow.
The X-ray flux of the source would imply a
distance greater than 30 kpc for near-Eddington luminosities. 
Magnetic fields capable of controlling the corona far from
the neutron star would also be able to control the bulk of
the accretion flow near the surface.
(The magnetic pressure scales as  $P_{\mathrm{B}} \propto B^2$
and for a dipole field $B^2 \propto r^{-6}$.
For higher-order components, the scaling is steeper.)
The absence of X-ray pulsations from such a system
can be explained by geometrical effects.
The influence of a magnetic field on the accretion flow could
introduce a non-Maxwellian electron distribution,
especially near the magnetic poles. 
This could in turn be seen as non-thermal hard
X-ray emission from the neutron star surface, which is
not observed. A corona controlled by the magnetic field 
could have a rather large vertical extent, as charged particles
in a magnetic field can move rather freely along the field-lines.
A mechanism that keeps the corona close to the disk so that the input
photons of the Comptonization are mainly from the soft disk
radiation and not from the boundary layer is hard to find.

A `corona' that is actually the upper part of the accretion disk,
could also explain the observed properties.
Such models usually have relatively low optical depths of the corona
\citep{Pou96}. The source of soft input photons
is very close to the Comptonizing electrons.
When viewed from the neutron star, the solid angle covered by the
Comptonizing disk portion will be quite low, so the absence of
processed boundary layer emission can be explained easily in this
scenario. An increase in the accretion rate is likely to
increase the vertical extent of the disk and the surface
layer temperature, due to increased viscous heat release.
An optically thick surface layer with a temperature in the keV
range would practically prevent cooling of the bulk
of the disk, resulting in mass loss, possibly through a disk wind.
The response of the relative thicknesses of the layers to
changes in accretion rates is unclear. Improved models
for the interdependence of accretion rate and vertical
disk structure are needed for a more quantitative discussion.
This `flared-disk' scenario seems to be an adequate explanation
for the observed X-ray spectra. 

\subsection{Iron line}

The flux of the Comptonized component and the flux of
the 6 keV \feka~ feature are both higher in the high state. 
In the low state, the line is undetected, with a flux
lower by a factor of at least five to ten. This correlation
suggests the components could be physically related.
Potential sources for the \feka~ feature are
collisional excitation \citep{Arn85}, 
radiative recombination \citep{Arn92},
fluorescent emission and Compton reflection \citep{Mag95}.

The parameters of the \feka~ line in the RXTE data are similar
to those of the EXOSAT observation \citep{Singh, Got95}.
The line energy is slightly
above the 6.4 keV value, indicating ionization states
up to $\ion{Fe}{xxv}$ \citep{Nag89}.
Collisional excitation would require
temperatures of a few $\, \mathrm{keV}$~ \citep{Arn85} to produce
a line where the high ionization states dominate.
These temperatures are similar to the
observed temperature of the Comptonized component.
The high line flux differences between the states 
(at least one order of magnitude) when the temperature drops
by a factor of 4--5 is not straightforward to explain
by pure collisional excitation.
Photoionization may also influence the ionization state,
as the gas emitting the \feka~ line is irradiated by
the accretion disk and the boundary layer.
Radiative recombination of photoionized iron may be partially
responsible for the line. More detailed modeling of the environment
and better observational constraints on
the \feka~ line are needed before any firm conclusions
regarding these mechanisms can be drawn.

If the line is produced by Compton reflection or
fluorescence, the line flux should be proportional
to the continuum providing the input energy or
`seed photons' for the line emission mechanism.  
For fluorescence, the seed photon energy is near
$8 \, \mathrm{keV}$ (iron absorption edge).
The flux ratio of the Comptonized component in this band
between the states is $\simeq 200$. As variations in
the ionization state are not likely to change the
fluorescent yield with a factor larger than $10$
\citep{Kal95}, fluorescence can not be ruled out by flux
considerations. However, the large line width is hard to
explain with fluorescence.

The Compton reflection seed photons have an energy very close
to the line energy. In the low state, the flux of the Comptonized
component ($6$--$7 \, \mathrm{keV}$) is lower by a factor 
of $\simeq 20$. The ratio of the line and continuum
($6$--$7 \, \mathrm{keV}$) fluxes in the high state is
$0.04$--$0.1$, suggesting a face-on disk \citep{Mag95}.
If the ratio of Comptonized and reflected components
remains constant during phase transitions,
the line flux should be well below the detection limit in
the low state. The fact that the BB component is not
absorbed by the thick corona supports the face-on disk hypothesis.
The above discussion shows that the mechanism producing the 
\feka~ feature detected in the RXTE data can not be deduced
from these observations alone.

\subsection{Absorption and the 0.67 keV feature}

The galactic coordinates of \targ~ are
$\beta = -6.3$ and $\lambda = 321.8$.
The total galactic hydrogen column density in this direction is
$N_{\mathrm{H}} \approx 30 \cdot 10^{24} \, \mathrm{m}^{-2}$ \citep{Dic90}.
The X-ray spectral fits give similar values for the $N_{\mathrm{H}}$,
suggesting that \targ~ is at a distance greater than 10 kpc.
Another possibility is that there is a local ISM component
related to \targ, increasing the absorption above the
galactic value. The $0.67 \, \mathrm{keV}$ feature could be an artifact
of the local ISM \citep{Jue01}, as interstellar absorption features
complicate the analysis of the  $0.67 \, \mathrm{keV}$
feature. The neon absorption edge is very close to this,
and the absorption of the ISM makes the continuum slope rather
steep at these energies (see Fig. \ref{ASCA}). 
These complications might also cause the changes in 
line/edge parameters when the continuum model is changed.
The small differences in line parameters
when comparing my results to those of Juett et al. (2001)
are likely to be due to the different continuum model
(powerlaw and BB with $kT \approx 0.4 \, \mathrm{keV}$).
Juett et al. (2001) suggest observations
with high spectral resolution near $0.67 \, \mathrm{keV}$
to distinguish between an enhanced absorption edge and an emission line.
Emission would more likely be related to only one of the
continuum components, as absorption affects the
total spectrum. As the ratio of continuum component fluxes
varies considerably, observations in the high state with energy
resolution comparable to ASCA or SAX would also help in
finding the cause of the $0.67 \, \mathrm{keV}$ feature.
Unfortunately all observations with instruments capable of detecting the 
the $0.67 \, \mathrm{keV}$ feature have been made in the low state.

The $N_{\mathrm{H}}$ from the X-ray spectral fits presented above
should give a visual extinction of $A_{\mathrm{V}} \approx 1.8$ 
\citep{Pre95, Boh78}. The optical counterpart
has been identified as a $B \approx 20$ magnitude star \citep{McC78}. 
The LMXB optical emission is usually dominated by the accretion disk. 
The intrinsic color indices are in the range 
$(B-V)_{\mathrm{0}} = [-0.5, 0.5]$, the average being 
around $-0.2$  \citep{VPa95}. 
The known absolute visual magnitudes are between 
$M_{\mathrm{V}} = [-5, 6]$ \citep{VPa94}, and the weaker sources with 
$M_{\mathrm{B}} < 0$ are generally X-ray bursters. 
After correcting for absorption, the ratio of
X-ray ($2$--$10 \, \mathrm{keV}$) to optical luminosity is
$L_{\mathrm{X}}/L_{\mathrm{opt}} \approx 480$, 
a typical LMXB value \citep{VPa95}.

\begin{figure}
\resizebox{\hsize}{!}{\psfig{figure=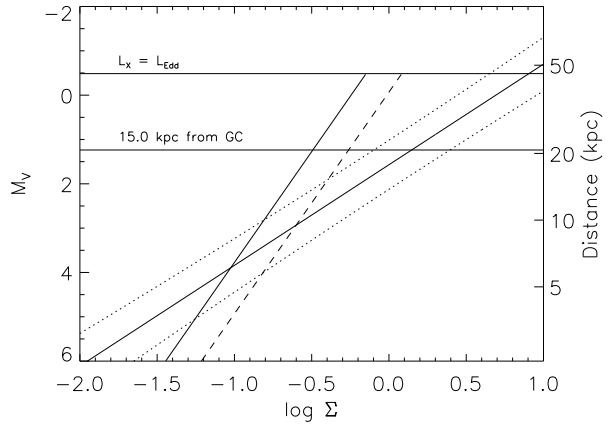,angle=0}}
\caption{Absolute visual magnitude $M_V$ (left axis) and
distance (right axis) of \targ~ against
$\Sigma = P_{\mathrm{hr}}^{2/3} (L_{\mathrm{X}}/L_{\mathrm{Edd}})^{1/2}$.
The solid and dotted lines indicate the relation
$M_{\mathrm{V}} = 1.57 - 2.27 \log \Sigma$ with error estimates \citep{VPa94}.
An additional error term of 0.5 mag representing inclination
variations is added to the original error estimates.
To estimate the V magnitude,
an intrinsic color $(B-V)_{\mathrm{0}} = -0.2$ is assumed.
The column density of X-ray measurements has been used to estimate
the extinction $A_{\mathrm{B}} = 1.8$ 
and color excess $E_{\mathrm{B-V}} = A_V/3 = 0.6$
The horizontal lines are upper limits for the distance
assuming $L_{\mathrm{X}} = L_{\mathrm{Edd}}$ 
(for a 1.5 $M_\odot$ NS and a He-rich donor)
and 15 kpc distance from the Galactic Center.
Note that the galactic latitude implies a distance of
$z = r \sin \beta \approx 0.11 r$ from the Galactic Plane.
Constant-period lines represented by solid (P = 36 min) and
dashed lines (P = 80 min) indicate the region of degenerate
donors. Systems with periods below 36 min and observable accretion
rates are most likely produced by tidal capture.  
It seems possible that the system has a very short period.}
\label{mv_logxi}
\end{figure}

The galactic coordinates of \targ~ ($\beta = -6.3$ and $\lambda = 321.8$)
imply a minimum distance between the source and Galactic Center of
$\approx 5.6 \, \mathrm{kpc}$ (assuming a distance of 8.6 kpc for the 
Galactic Center).
Therefore \targ~ is probably a member of the disk population,
and the common literature value of distance $10 \, \mathrm{kpc}$,
more representative for the bulge population,
should be regarded as an order-of-magnitude estimate.
Using the empirical relation between optical and X-ray fluxes and
the period \citep{VPa94}, it can be concluded that the system may
have a degenerate donor (Fig. \ref{mv_logxi}),
and the distance is likely to be in the range
$3$--$20 \, \mathrm{kpc}$.

\section{Conclusions}

Archival X-ray observations of \targ~ have been analyzed.
The X-ray continuum can be fitted with a two-component model consisting
of an isothermal blackbody and a Comptonized component.
Two different X-ray states are seen. In the high state, the luminosity
comes mainly from the Comptonized component and the spectrum is
harder. The low state spectrum is dominated by the BB,
but the Comptonized component is also important at
energies below  $2 \, \mathrm{keV}$. 
Two gaussian features at $6.5 \, \mathrm{keV}$ (the \feka ~ line) and
$0.67 \, \mathrm{keV}$ are detected. The \feka~ line
is seen only in the high state, and it has at least 
one order of magnitude lower flux in the low state.
A two-layer disk, with
the lower and cooler layer providing the input photons
to be upscattered by the hotter surface layer,
provides a qualitative explanation for the X-ray continuum
and state transitions. 

The BB component of the X-ray spectrum 
can be taken as weak evidence of a neutron star. It is probably
safe to assume that the BB comes from a boundary
layer between the neutron star surface and the inner disk,
and the Comptonized component is from a hot disk corona.
The ratio of the continuum component luminosities is close to unity.
This is in good accordance with the theoretical expectation for
a slowly rotating neutron star \citep{Sun86}. Most LMXBs
have lower BB luminosities \citep{Whi88}.
A rapidly rotating neutron star or non-thermal emission from
the boundary layer have been suggested to explain this discrepancy.
In LMXBs with detected BB emission, the
BB luminosity correlates strongly with total luminosity.
The BB luminosity of \targ~ decreases slightly
when the total luminosity increases significantly.
The changes seen in continuum spectrum and its variations,
which are oppposite to observations of other LMXBs,  
can be explained if \targ~ has
a boundary layer with a spectrum close to BB and 
a slowly rotating neutron star ($ P \gg \, \mathrm{ms}$).

Angular momentum of the accreted matter tends to spin up the
neutron stars. An equilibrium between angular momentum gain
from the accretion flow and losses due to gravitational radiation
is reached at periods of the order one millisecond. 
For typical LMXB accretion rates the equilibrium is reached
in a time of the order $10^7$ years \citep{Rap83}.
The observations suggest that \targ~ has gained less angluar momentum.
It has either been accreting for a shorter time than a typical LMXB,
or the accretion rate is significantly smaller.
Monitoring observations with sufficient energy resolution to estimate
the continuum components would be especially useful.
Such observations have the best possibility of observing
the system in a wide range of states.

The gaussian feature seen in the RXTE high state spectra 
near $6.5 \, \mathrm{keV}$ is interpreted as the \feka ~ line.
The line parameters are partially produced by instrumental
effects, and should be treated with some caution.
As the line is absent in the low state,
with an upper flux limit one order of
magnitude lower than the observed RXTE flux, the line
is probably related to the Comptonized component.
The line is broad and the centroid energy is above
the neutral iron value, suggesting it might originate from
ionized gas. Possible mechanisms producing this line are radiative
recombination \citep{Arn92}, collisional excitation \citep{Arn85}
and Compton reflection \citep{Mag95}. The line width implies
fluorescence is less likely to be responsible for the line.
High S/N low-state spectra are needed to provide better constraints
on the line formation. If the low-state \feka ~ fluxes are close
to the upper limit derived from ASCA data, more detailed modeling
than presented here is needed. On the other hand, very low
line fluxes would favour the Compton reflection mechanism.
The two-layer disk model providing the best qualitative explanation
for the Comptonized spectrum could easily produce such
a reflected component.

The $0.67 \, \mathrm{keV}$ feature has been detected previously in
ASCA data \citep{Jue01}. It also detected in SAX data,
and the parameters do not change significantly, when a different
continuum model is used. This analysis confirms the detection of the
$0.67 \, \mathrm{keV}$ feature.
The feature is either a neon absorption
edge, made stronger by enhanced neon abundace \citep{Jue01},
or an emission line. Juett et al. (2001) suggest
observations with very high spectral resolution
to distinguish between the two mechanisms producing the feature.
The detected state transitions show that the distinction could also
be made by comparing medium-resolution observations in the
two states. If the feature strength changes with state, the
line interpretation is more likely. Changes in the continuum
component physically related to the line are likely to produce
changes in the line parameters. If the equivalent width of the
feature remains unchanged during state transitions, it is more
likely to be the absorption edge.

Optical spectroscopy of the system would allow determining the
abundances of the inflowing matter. The equivalent
widths of the lines near $460 \, \mathrm{nm}$
$\ion{He}{II}$, $\ion{C}{III}$, $\ion{N}{III}$ and
H$\beta$ could provide the needed abundance diagnostics.
A short-period system with a degenerate donor would
have stronger He and possibly CNO lines and weaker
H$\beta$ line than a system with a main sequence donor.

\begin{acknowledgements}
I am grateful to Pasi Hakala, Panu Muhli
and Osmi Vilhu for useful discussions, and to Diana Hannikainen for
both useful discussions and checking the English of the manuscript. 
I thank the anonymous referee for his/her useful comments. 
This research has made use of NASA's Astrophysics Data System (ADS)
Bibliographic Services and data obtained from the High Energy
Astrophysics Science Archive Research Center (HEASARC), provided
by NASA's Goddard Space Flight Center, and
the SIMBAD database, operated at CDS, Strasbourg, France.
Financial support of Academy of Finland and the National
Technology Agency TEKES is acknowledged.
\end{acknowledgements}
\bibliographystyle{aa}

\end{document}